\documentclass[superscriptaddress,twocolumn,showpacs,prb]{revtex4-1}
\usepackage[utf8]{inputenc}
\usepackage{amsmath}
\usepackage{braket}
\usepackage{graphicx}
\usepackage{amsfonts}
\usepackage{pgfplots}
\usepackage{csquotes}
\usepackage{hhline}
\usepackage{amssymb}
\usepackage{listings}
\usepackage{color}
\definecolor{codegreen}{rgb}{0,0.6,0}
\definecolor{codegray}{rgb}{0.5,0.5,0.5}
\definecolor{codepurple}{rgb}{0.58,0,0.82}
\definecolor{backcolour}{rgb}{0.95,0.95,0.92}
 
\lstdefinestyle{mystyle}{
    backgroundcolor=\color{backcolour},   
    commentstyle=\color{codegreen},
    keywordstyle=\color{magenta},
    numberstyle=\tiny\color{codegray},
    stringstyle=\color{codepurple},
    basicstyle=\footnotesize,
    breakatwhitespace=false,         
    breaklines=true,                 
    captionpos=b,                    
    keepspaces=true,                 
    numbers=left,                    
    numbersep=5pt,                  
    showspaces=false,                
    showstringspaces=false,
    showtabs=false,                  
    tabsize=2
}

\usepackage{subfigure}

\usepackage{dcolumn}
\usepackage[qm]{qcircuit}
\usepackage{tabularx}
\usepackage[colorlinks=true,linkcolor=blue,citecolor=blue,urlcolor=blue]{hyperref}
\usepackage{longtable}
\usepackage{braket}
\usepackage{float}
\newcolumntype{C}{>{\centering\arraybackslash}X}
\begin{document}
\title{Cancer Detection Using Quantum Neural Networks: A Demonstration on a Quantum Computer}

\author{Nilima Mishra}
\email{b217024@iiit-bh.ac.in}
\affiliation{International Institute of Information Technology Bhubaneswar, Gothapatna, Bhubaneswar, 751003, Odisha, India}

\author{Aradh Bisarya}
\email{ph1170810@physics.iitd.ac.in}
\affiliation{Indian Institute of Technology Delhi, New Delhi, India}

\author{Shubham Kumar}
\email{shubhamkumar.kumar@gmail.com, shubham@shu.edu.cn}
\affiliation{International Center of Quantum Artificial Intelligence for Science and Technology~(QuArtist) and Physics Department, Shanghai University, 200444 Shanghai, China}

\author{Bikash K. Behera}
\email{bkb18rs025@iiserkol.ac.in}
\affiliation{Department of Physical Sciences,\\ Indian Institute of Science Education and Research Kolkata, Mohanpur 741246, West Bengal, India}

\author{Sabyasachi Mukhopadhyay}
\email{mukhopadhyaysabyasachi@gmail.com}
\affiliation{BIMS Kolkata, Kolkata- 700 097, West Bengal, India}

\author{Prasanta K. Panigrahi}
\email{pprasanta@iiserkol.ac.in}
\affiliation{Department of Physical Sciences,\\ Indian Institute of Science Education and Research Kolkata, Mohanpur 741246, West Bengal, India}

\begin{abstract}
Artificial intelligence and machine learning paves the way to achieve greater technical feats. In this endeavor to hone these techniques, quantum machine learning is budding to serve as an important tool. Using the techniques of deep learning and supervised learning in the quantum framework, we are able to propose a quantum neural network and showcase its implementation. We consider the application of cancer detection to demonstrate the working of our quantum neural network. Our focus is to train the network of ten qubits in a way so that it can learn the label of the given data set and optimize the circuit parameters to obtain the minimum error. Thus, through the use of many algorithms, we are able to give an idea of how a quantum neural network can function.
\end{abstract}


\maketitle

\section{Introduction}
Techniques to mimic human intelligence using logic, algorithms, and networks are the state of the art findings of science and technology. Artificial intelligence with its subsets of machine learning and deep learning takes a giant leap in analyzing and classifying data. The neural network proves to be the perfect programming-paradigm in this field \cite{qnn_AspuruArxiv2017,qnn_Bajoninpj2019,qnn_AjitInfSci2000}. With deep learning as its backbone, these neural networks help in areas of speech recognition, image recognition, data analysis, and many more.

Deep learning employs the idea of managing the input data set through several layers of representation \cite{qnn_HintonNat2015}. It uses each layer to learn the information deeply. The learning can be supervised or unsupervised. In supervised learning, we try to learn the function from the input training set. In this learning method, each input training set is labeled with the desired output value. Thus, during the training phase, the system has an idea of what the output should be like for the given input. In unsupervised learning, the data is analyzed without it's labels so it is not known what the output should look like. It is mostly used to understand the data better. This entire process of deep learning is inspired by the functioning of the human brain and thus we implement it through the neural networks (artificial neural network).

Amalgamating artificial neural network (ANN) models with concepts of quantum information and quantum mechanics, we are able to propose a quantum neural network (QNN) model. The advantage of training the neural networks with big data analysis, with features like quantum computing gets an edge over artificial neural network by quantum neural network. However, quantum neural network models are mostly theoretical proposals as their full implementation in the physical world is yet to come.

In this paper, we attempt to build a quantum neural network model, exhibiting the use of several algorithms to handle the data set. Out of many applications of neural networks, we have taken cancer detection into account \cite{qnn_JainJCSSB}. An attempt to train the network to calculate the label function and minimize the loss function using standard algorithms has been done. We have tried to detect cancer on the basis of the label found from the input data set. In the first place, we tried to give an insight into the process of detection of the disease by the image recognition method. In the second place, we tried to implement several algorithms on a numerical data set consisting of parameters associated with the cell size and texture, as our input data set. The results obtained are quite convincing demonstrating the application of the neural network. This work is a step taken to showcase one of the uses of a quantum neural network, which would definitely encourage to explore more into the field.

IBM quantum experience, due to its easy and free access to the cloud-based quantum computer, has become a large platform for the research community in the field of quantum computation and quantum information to accomplish various tasks such as quantum simulation \cite{qnn_KapilarXiv:1807.00521,qnn_ManabputraarXiv2018,qnn_MalikRG2019,qnn_KlcoPRA2018,qnn_SonkarRG2019,qnn_MalikarXiv2019,qnn_ZhukovQIP2018}, quantum algorithms \cite{qnn_GangopadhyayQIP2017,qnn_DasharXiv2017,qnn_SrinivasanarXiv:1801.00778}, quantum information-theoretical demonstrations \cite{qnn_KalraQIP2019,qnn_GhosharXiv2019,qnn_SwainQIP2019,qnn_GuptaRG2019}, quantum machine learning \cite{qnn_MishraRG2019}, quantum error correction \cite{qnn_HarperPRL2019,qnn_NishioarXiv2019,qnn_SingharXiv2018,qnn_AshutoshRG2019}, quantum applications \cite{qnn_BeheraQIP2017,qnn_BeheraQIP2019,qnn_BeheraQIP2019} to name a few. Here, we use this platform for the task of implementing QNN as an application to cancer detection.

\begin{figure*}
    \centering
    \includegraphics[width=\textwidth]{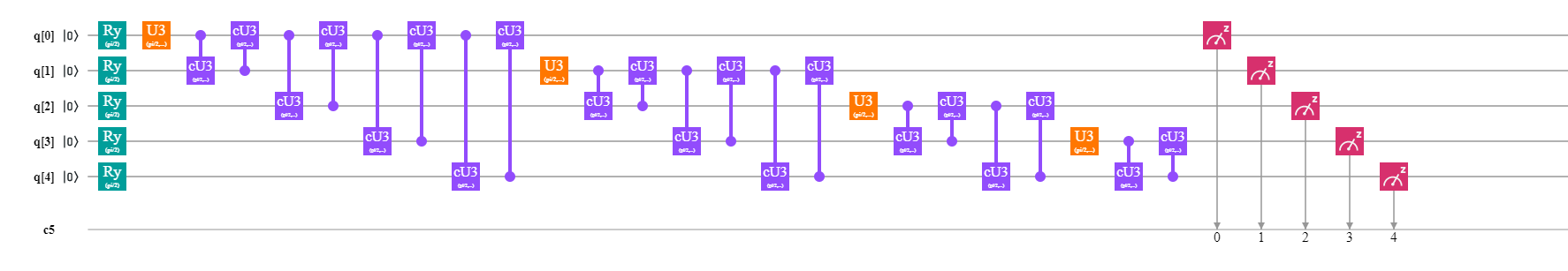}
    \caption{The maximally entangled qubit network is represented on IBMQ. This system scales up as O(${N^2}$) with the number of qubits.}
    \label{qnn_img1}
\end{figure*}

\begin{figure}[]
    \centering
    \includegraphics[scale=0.27]{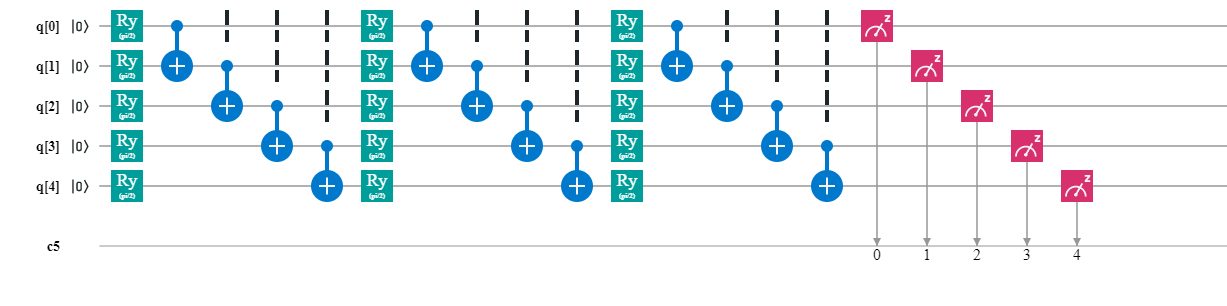}
    \caption{IBMQ system demonstrating how partial entanglement can be used to reduce the number of parameters required in a quantum neural network to O($N$).}
    \label{qnn_img2}
\end{figure}

\begin{figure}
    \centering
    \includegraphics[scale=0.18]{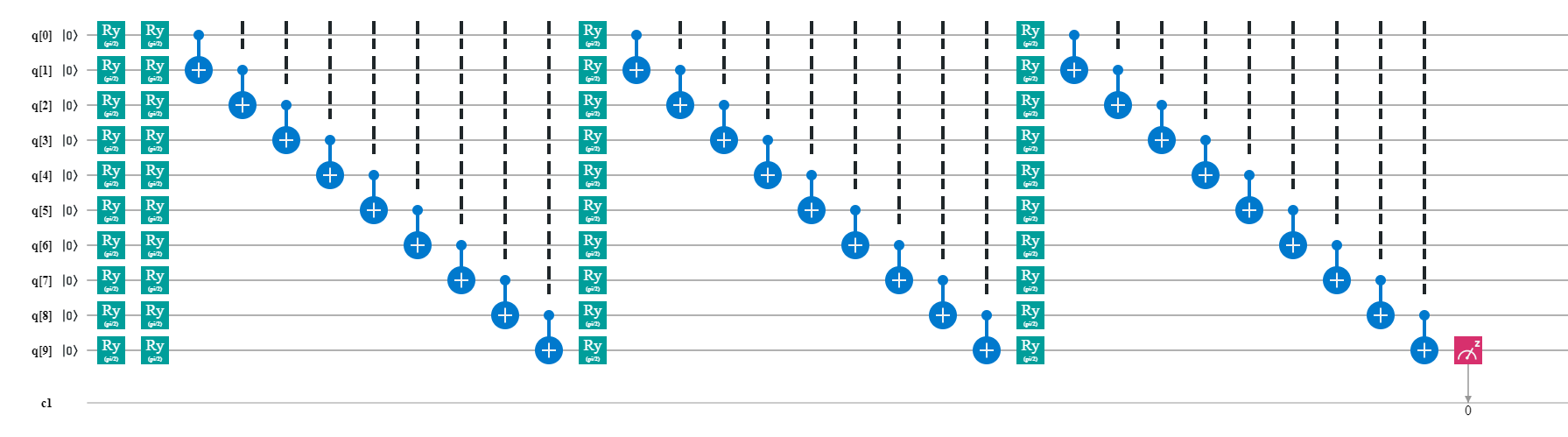}
    \caption{10 qubits partially entangled quantum neural network.}
    \label{qnn_img3}
\end{figure}

\begin{figure}
\centering
\includegraphics[width=0.5\textwidth]{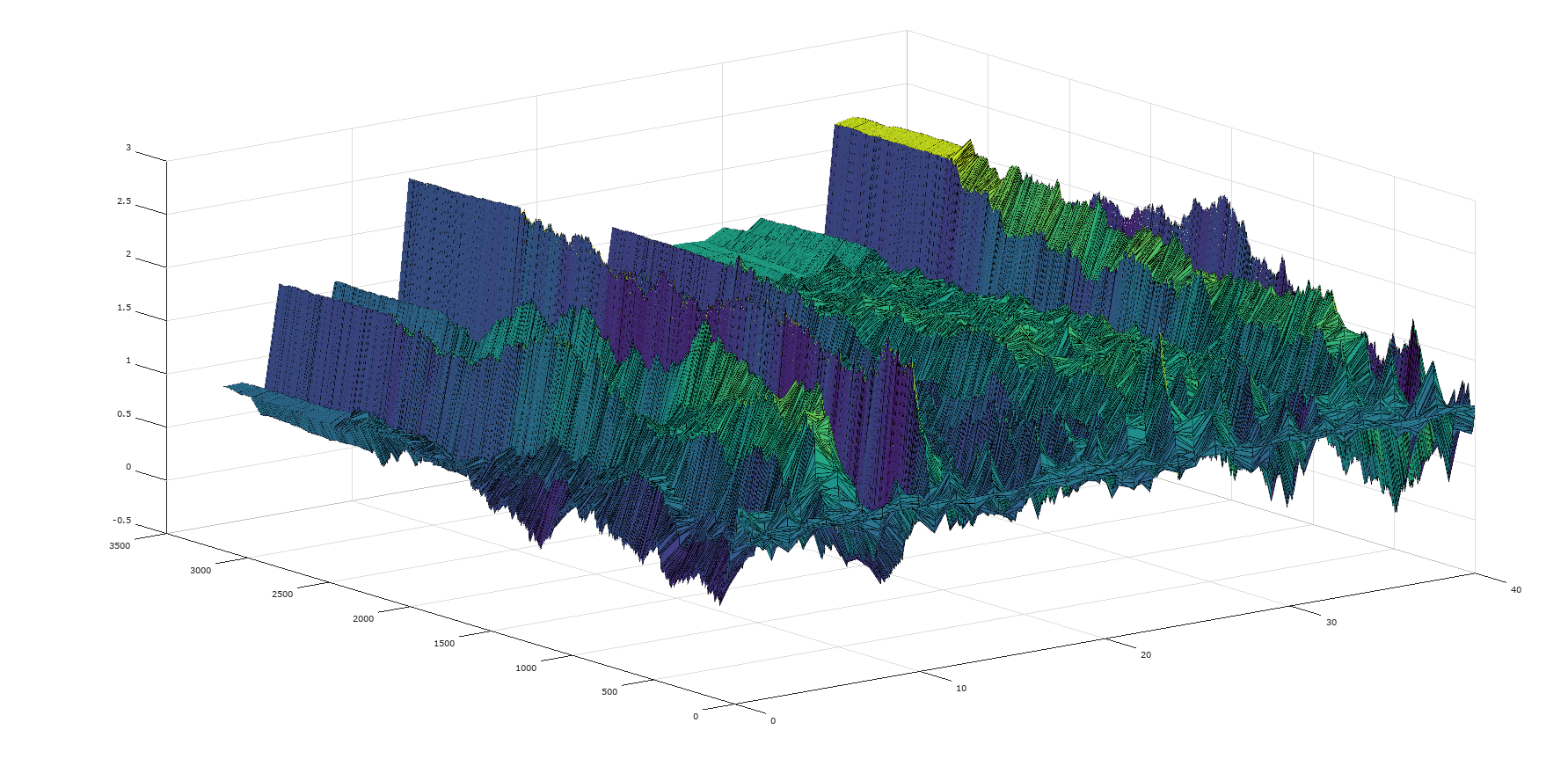}
\caption{\textbf{Variation in the parameters over number of iterations}.}
\label{qnn_img4}
\end{figure}

\begin{figure}
\centering
\includegraphics[scale=0.3]{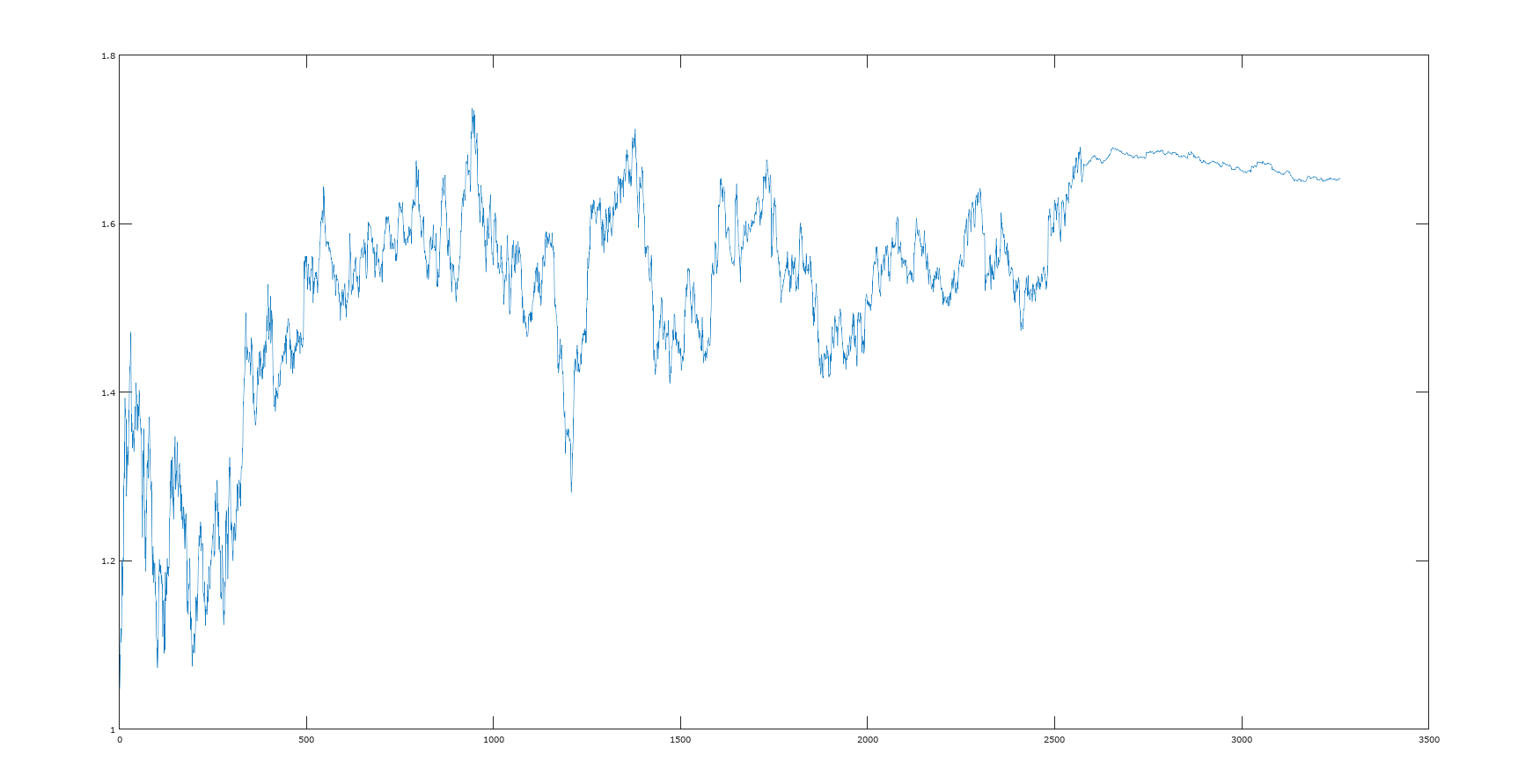}
\caption{\textbf{Representative variations of a parameter. Easily visible is the definite upward trend, and the point where the learning rate is manually decreased}.}
\label{qnn_img5}
\end{figure}

\begin{figure}[]
\includegraphics[scale=0.23]{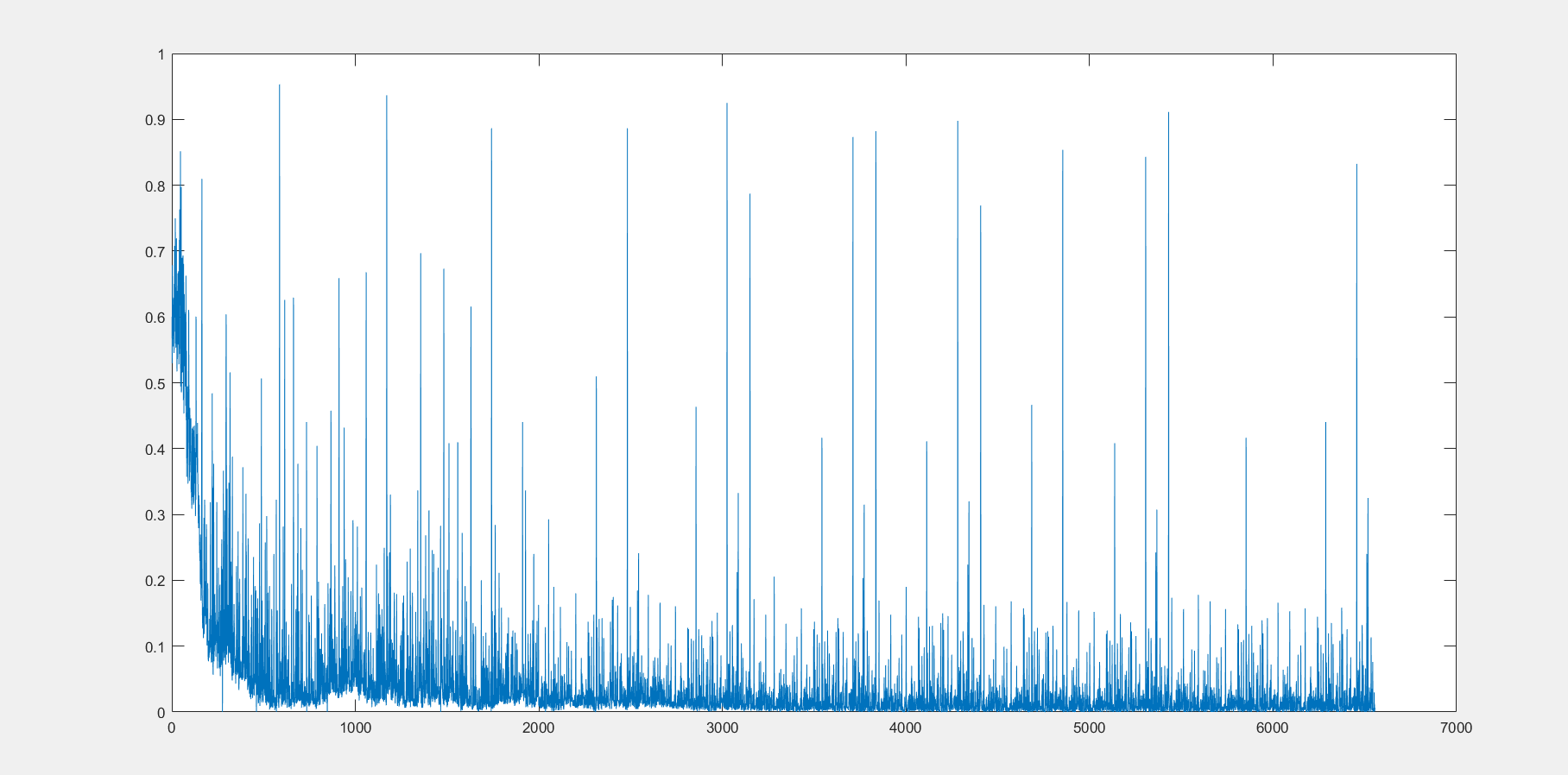}
\caption{\textbf{The value of the loss function over the number of iterations. The peaks correspond to anomalous data, but the overall decrease is clearly visible}.}
\label{qnn_img6}
\end{figure}

\begin{figure}[h]
    \centering
    \includegraphics[scale=0.3]{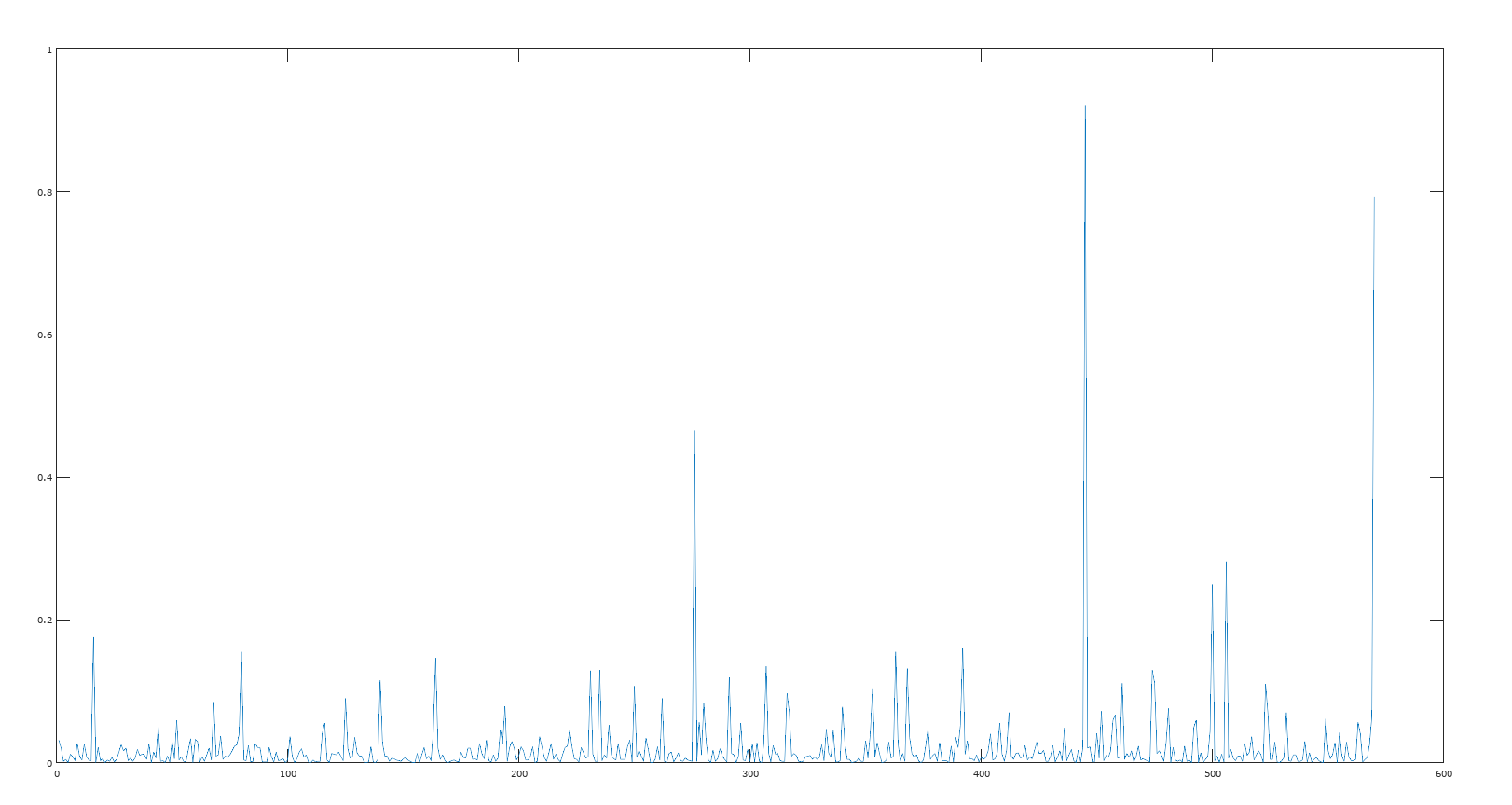}
    \caption{Loss over all inputs for training done over the entire set.}
    \label{qnn_img7}
\end{figure}

\begin{figure}[]
    \centering
    \includegraphics[scale=0.3]{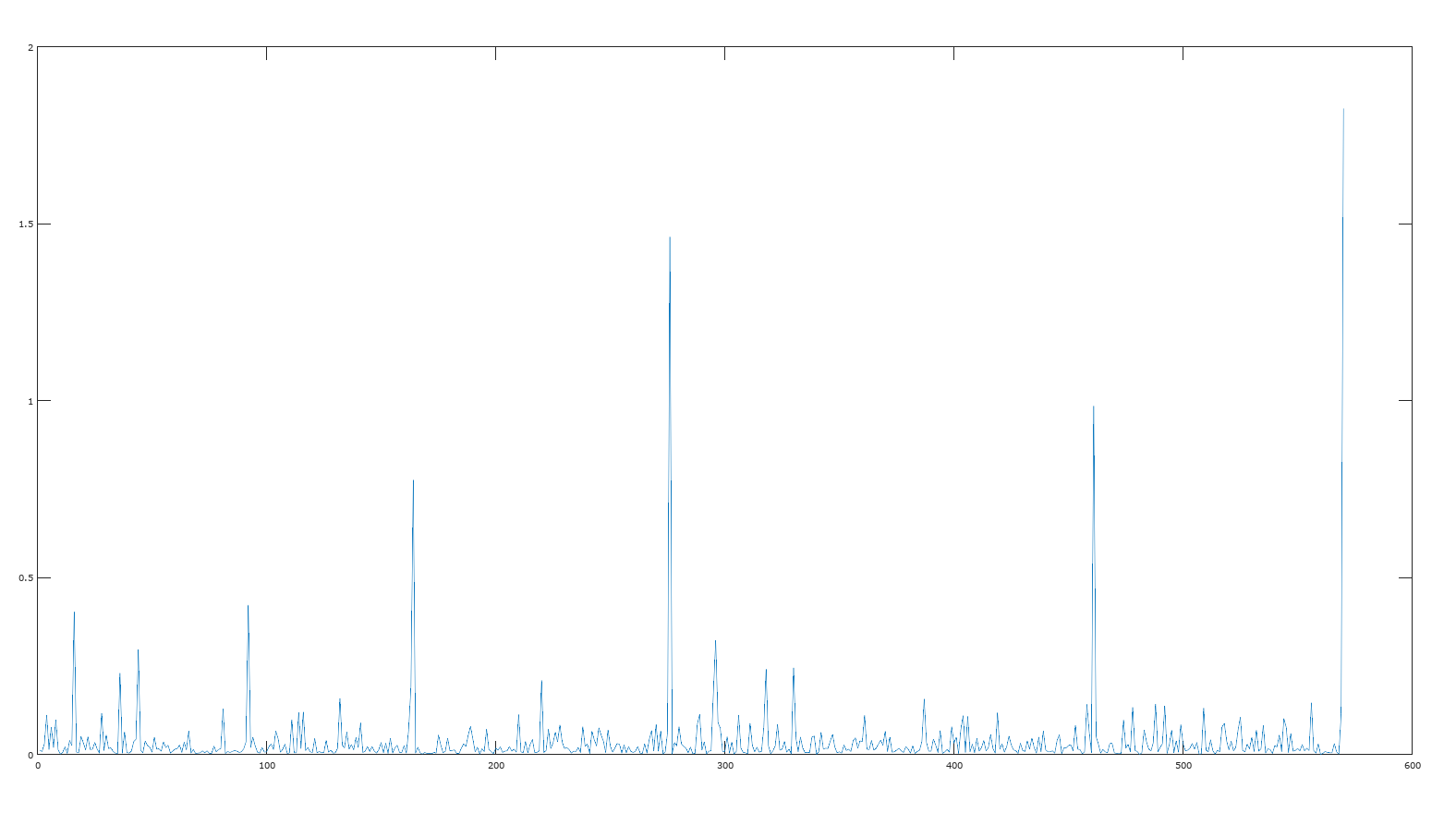}
    \caption{Loss over all inputs for training done over the first 100 elements.}
    \label{qnn_img8}
\end{figure}

\section{methods \label{qnn_Sec2}}
Deep learning along with supervised and unsupervised learning have given an edge over other methods in handling data. The product of these techniques is the architecture of a neural network. These artificial neural networks, through their layered structure, help the system to learn data handling. The input set of data, usually known as the training data, comes with some label value. It is then processed through the layers of the network such that the output label obtained is close to the true label. This basic framework is then extended to applications of image and pattern recognition, text classification, speech recognition and many more.

Taking inspiration from the design and functioning of the classical neural network, we have designed a quantum neural network, capable of operating on a 10-qubit system. The quantum neural network designed is capable of performing several algorithms to optimize functions or calculate the label of strings. Our design helps in obtaining the training vector to calculate the label value of the string given as input. Demonstrating the implementation of the network, we have tried to use the basic principles of machine learning to handle data, which we have demonstrated through the application of detecting cancer. We tried with two methods, first by training through the numerical dataset and also tried to utilize the method of image recognition. We carried out all the implementation and simulation on the cloud-based platform provided by IBM.

\subsection{Using the data set consisting of information on cell features}

\begin{center}
We detect whether the person suffers from cancer using the biodata providing information on the size, radius, etc., of the affected cell, by implementing a quantum neural network.
\end{center}

\textbf{Background:}
Classification of data has been an important part of machine learning. The principles of data classification and deep learning can aid us to detect cancer in a patient. We, thus, design a quantum neural network and demonstrate its implementation by detecting breast cancer in the patient.

\textbf{The quantum setting of the process:}
Analogous to neurons and weights in a classical neural network, we have qubits and their connections along with the rotation gates. The design is carried on a 10-qubit system. Our main intention is to obtain the training parameter that implies to minimum deviation of the obtained label value from its true value. Each string as an input has some label value assigned to it. The inputs are fed to the circuit having some rotation parameters. The training vector is varied after every iteration to correspond to minimum loss. The optimization algorithm of online gradient descent has been implemented. The rate of decrease is also regulating. This method gives successful interpretations on the operation of the network giving further results on the analysis of variation of training parameters and loss functions. The circuit design and implementation is carried through the IBM quantum experience.

\subsection{Process of Image Recognition}

\begin{center}
We demonstrate how to design the quantum neural network that can be used to detect the disease through the image recognition process.
\end{center}

\textbf{Background:}
The process of detection of the disease can also be carried out through the method of image recognition. The image is broken into an array consisting of values basing on its brightness.

\textbf{The quantum setting of the process:}
The process of image recognition is basically basing on the grey scale principle. The system assigns values to each pixel, ranging from 0 to 1 based on its greyscale, that is, how dark it is. These values are then fed as the input training set in the form of an array. The network is then trained to find the label value of the input set and the error is calculated based on its deviation from the true label. We collect the data set, i.e., the images from Kaggle, wherein we got the MRI images of the cells. The process involves entering the pixel matrix as the input to the neural network. However, the size of the matrix owing to the image is too large to be given as input to the currently available quantum simulator circuit. Therefore, we chose a $4\times4$ matrix of pixels from the image and enter it as the dataset. The training parameter is thus calculated which gives the minimum error or loss. To minimize the loss function, we implement the optimization algorithm of a Variational Quantum Eigensolver. We determine to find the training parameter which gives the minimum error while calculating the label function. The parameters are thus chosen which are optimal to the given problem.

The matrix that is sent as an input is selected from the image. A validation set is obtained which is an array of size $4\times4$, taken randomly from the image. This is then coded into the network for classification and to obtain the label value of the training set. However, the entire 256 qubit system could not be simulated owing to the unavailability of the required system. We act on the input set using rotation gates of Z gates and X gates to gradually change the training parameter following optimization to give the minimum deviation of the obtained label value from the true label value. The simulation is carried using the API based cloud access provided by IBM. The method, however, gives down-sampled results owing to the random selection of apart from the image and not sending the entire picture, due to lack of required system. Though the use algorithm can be leveled upon the availability of a higher qubit handling a quantum computer. 

\section{Architecture \label{qnn_Sec3}}
Neural networks may be implemented in quantum computers by treating the qubits as neurons. We act on neurons with rotation gates which have parameters, which are then optimized in our model. The inputs are implemented as initial rotations on each qubit. A CNN with $N_{i}$ neurons in layer I and m layers have parameters, one for each connection. This goes up as O(${N^2}$). This is where a QNN provides an advantage. If we were to fully incorporate the effect of each qubit on the others, we would still have an O(${N^2}$) model. This is illustrated for 5 qubits in Fig. \ref{qnn_img1}. The input is 5 dimensional, each input affecting the Y rotation parameter, and the output is a 5-bit number.

Each U3 has 3 parameters, which totals to a 72-dimensional parameter space. A CNN with biases have 60 dimension parameter space. In a CNN, dropping the number of parameters results in a lack of connections between neurons, however, we can implement partial connections by entanglement in a QNN using CX gates. Non parameterized gates allow us to entangle the inputs with lower parameter spaces. One such architecture is shown in Fig. \ref{qnn_img2}. A two layer architecture only requires 10 parameters and the number of parameters go up as O($N$).

\section{Cancer Detection \label{qnn_Sec4}}
We attempt to use our simplified QNN to detect breast cancer given patient biodata. In theory, a simpler set such as the MNIST handwritten numerals or cancer cell imaging maybe use. However, 256 qubit systems cannot be simulated efficiently on currently available systems, and the exercise reduces to one of the theoretical interest for QNNs. The biodata (Wisconsin Breast Cancer Dataset) consists of features of lumps on the patient's breast, e.g., radius, roughness, concavity, etc. We take 10 parameters and assign each to one qubit. The initial Y rotations are set up so that $|0\rangle$ corresponds to the minimum value of the variable and $|1\rangle$ to the maximum. Also, since we only want to determine whether a lump is cancerous or not, we need one output bit so we only measure the last qubit's value. We assign 0 to no cancer (benign) and 1 to cancer present (malignant). The loss function is logarithmic. If l' is the obtained label, and l is the expected label,

\begin{equation}
loss=-(1-l)\cdot log(1-l) - l\cdot log(l)
\end{equation}

We then compute by varying each parameter one at a time. The parameters are then moved in the direction opposite the gradient, controlled by a learning rate r. As the differences in values become smaller, r is also set to smaller values.

\begin{equation}
\vec{p'}=\vec{p}-\frac{r\cdot\vec{\nabla}(loss)}{\vec{\nabla}(loss)}
\end{equation}

The data set is of size 600. We first take the whole data set as our training set, and then sample a subset for training, to treat the remaining as test data.

\section{Results and Discussion \label{qnn_Sec5}.}
The variation of the parameters over time is shown below.

Below is a representative parameter. The point where the learning rate is manually changed can be observed.

The loss over most inputs slowly decreased. Some anomalous inputs resulted in wrong answers and excessive losses.

The final losses for all 600 inputs are shown below. The average loss is 2.1\%. 6.67\% of the inputs and have losses greater than 1\%.

The process is also carried out with a training set of 100, and the average loss over the entire set is 3.9\%. 6.67\% inputs have an error greater than 5\%. Most of these lie in the 10-20\% range, similar to the previous case, which shows that there is a higher chance they exist due to input data inconsistencies, and less due to error in the final parameters. For all purposes, the network is sufficiently trained in cancer detection.

\section{Conclusion \label{qnn_discussion}}
To conclude, we have demonstrated here the cancer cell detection using quantum neural networks to detect breast cancer. Using the proposed algorithms, we run the circuit for different parameteres and optimize it for the marked cell as the cancer cell. Using the techniques of deep learning and supervised learning in the quantum framework, we have proposed the quantum neural network and illustrated its implementation on the quantum simulator available at IBM quantum experience platform. We have considered the application of cancer detection to explicate the working of our quantum neural network. We have trained the network of ten qubits in a such way that it learns the label of the given data set and optimizes the circuit parameters to obtain the minimum error.

\section{Acknowledgements \label{qnn_acknowledgements}}
N.M., A.B., and S.K. acknowledge Indian Institute of Science Education and Research Kolkata for providing hospitality during the course of the project work. B.K.B. acknowledges the financial support of IISER-K Institute Fellowship. The authors acknowledge the support of IBM Quantum Experience for producing the experimental results. The views expressed are those of the authors and do not reflect the official policy or position of IBM or the IBM Quantum Experience team.

\clearpage
\iffalse

\section{Supplementary Information: Demonstration of a general fault-tolerant quantum error detection code for $(2n+1)$-qubit entangled state on IBM 16-qubit quantum computer}For simulating the error detection protocol, we used QISKit to take both simulation results. The QASM code for the same is as follows: 

\subsection{Measurement data}
We performed all the simulations on QISKit and recorded the countings of each of the measurement result over the two ancillary error syndrome qubit in 8192 shots. From the countings, the probability of each error \textit{i.e.} bit-flip error, phase-flip error, and arbitrary phase-change error was extracted. The data is shown in table \ref{qnn_table4} below.
\fi
\end{document}